\documentclass[twoside,11pt]{article}
\usepackage{amssymb}
\bibliography{bb}
\begin{document}
\title{\bf{Singleton physics\footnote
{This review is dedicated to our friend of 32.5 years, Ludwig Faddeev,
upon the occasion of his 65$^{\mathrm{th}}$ birthday. It was initiated
by the first author; the frame and a first draft were discussed between 
all three authors and the final formulation was completed by the latter
two after the sudden and untimely death of Mosh\'e Flato. To be published in 
a special issue of {\it Proceedings of Steklov Mathematical Institute}}}}
\author{{\sc{Mosh\'e Flato$^1$, Christian Fr\o nsdal$^2$ and 
Daniel Sternheimer$^1$}}\\
{\small {\it{$^1$ Laboratoire Gevrey de Math\'ematique Physique, 
CNRS \ ESA 5029}}}\\
{\small {\it{D\'epartement de Math\'ematiques, Universit\'e de Bourgogne}}}\\
{\small {\it{BP 400, F-21011 Dijon Cedex, France.}}}\\
{\small {{\rm e-mail:} {\tt flato@u-bourgogne.fr}, \
{\tt dastern@u-bourgogne.fr}}}\\
{\small {\it {$^2$ Department of Physics, University of California,}}}\\
{\small {\it Los Angeles, CA 90024-1547, USA}. \ {\rm e-mail:}
 {\tt fronsdal@physics.ucla.edu}}}
\date{}
\maketitle

\begin{abstract}
We review the developments in the past twenty years (which are based on
our deformation philosophy of physical theories) dealing with elementary
particles composed of singletons in anti De Sitter space-time. The study
starts with the kinematical aspects (especially for massless particles)
and extends to the beginning of a field theory of composite elementary
particles and its relations with conformal field theory (including very
recent developments).
\end{abstract}
\thispagestyle{empty}

\section{Introduction: Deformations and Singleton Physics}

Physical theories have their domain of applicability mainly depending on
the velocities and distances concerned. But the passage from one domain
(of velocities and distances) to another one does not appear in an
uncontrolled way. Rather, a new fundamental constant enters the modified
formalism and the attached structures (symmetries, observables, states,
etc.) {\it deform} \cite{Fl82,Fl98} the initial structure; namely, we have
a new structure which in the limit when the new parameter goes to zero
coincides with the old formalism. In other words, to {\it detect} new
formalisms we have to study deformations of the algebraic structures
attached to a given formalism.

The only question is in which category we perform this search for
deformations. Usually physics is rather conservative and if we start e.g.
with the category of associative or Lie algebras, we tend to deform in
this category. This is the case of traditional quantization
\cite{BFFLS78,St98} (deforming classical mechanics to quantum mechanics by
introducing a new parameter $\hbar$, keeping the same algebra of observables
but deforming their composition law). The same is true of the passage from
Galilean physics to special relativity (new parameter $c^{-1}$, where $c$ is
the speed of light) and thence to physics in De Sitter space-time
(the new parameter being the curvature). It is this last aspect which we
shall present here.

In this paper we touch recent developments in field theories
based on supergravity, conformal field theories, compactification of
higher dimensional field theories, string theory, M-theory, $p$-branes,
etc. for which people rediscovered the efficiency and advantages of
anti De Sitter theories (which are stable deformations of Poincar\'e field
theories in the category of Lie groups; see however \cite{FHT93} for
quantum groups, at roots of unity in that case). There are many reasons
for the advantages of anti De Sitter (often abbreviated as AdS) theories
among which we can mention that AdS field theory admits an invariant
natural infrared regularization of the theories in question and that the
kinematical spectra (angular momentum and energy) are naturally discrete.
But in addition AdS theories have a great bonus: the existence of
{\it singleton} representations discovered by Dirac \cite{Di63} for
${\rm{SO}}(2,3)$, corresponding to a ``square root" of AdS massless
representations. We discovered that fact around 20 years ago \cite{FF78,FF80}
and developed rather extensively its physical consequences in the following
years \cite{AF78,AFFS81,BFFH85,BFFS82,BFH83,FaF80,FF81,FF84,FF86,FF86f,%
FF87,FF88,FF89,FF91,FF98,FFG86,FFS88,FHT93,Fr79,Fr82,Fr88,FH87,HFF92}.

Singleton theories are topological in the sense that the corresponding
singleton field theories live naturally on the boundary at infinity of the
De Sitter bulk (boundary which has one dimension less than the bulk).
They are new types of gauge theories which in addition permit to consider
massless particles, e.g. the photon, as {\it dynamically} composite
AdS particles \cite{FF88,FF98,FFS88}. Some of the beautiful properties of
singleton theories can be extended to higher dimensions, and this is the
main point of the recent huge interest in these AdS theories, which touched
a large variety of aspects of AdS physics. More explicitly, in several of
the recent articles among which we can mention
\cite{Ma97,Wi98,FFr98,FFZ98,FZ98,FKPZ,FMMR,SS98},
the new picture permits to study duality between CFT on the boundary at
infinity and the corresponding AdS theory in the bulk. That duality, which
has also interesting dynamical aspects in it, utilizes among other things
the great notational simplifications permitted by singleton physics.

\section{Kinematics: one massless particle $=$ two Dirac singletons}
In order to give a flavor of the basic features of the theory of singletons
in the (2+3) anti-De~Sitter space-time AdS$_4$ and their relation to
massless particles, we shall in this section and in the following
indicate some of these features, refering to the quoted literature for
a more detailed presentation. The theory can be extended to other dimensions
(higher or lower). In AdS$_3$ one gets essentially the same features
\cite{HFF92}; the main difference being (as is well known) that the ($2+2$)
De Sitter group ${\rm{SO}}(2,2)$ is not the full (infinite-dimensional)
conformal group of $1+1$ space-time (one has then to study Witt and
Virasoro algebras; cf. e.g. \cite{BH86,Mi99,It98}). In space-times of
dimension $\geq 5$, some care is needed \cite{AL98,La98} as to the definition
of masslessness and of singletons (the very nice properties of dimension~4
are not all preserved, a fact sometimes overlooked in the recent
literature); we shall not enter here into this discussion.

The maximal compact subalgebra of $\mathfrak{so}(2,3)$ is
$\mathfrak{so}(2)\oplus \mathfrak{so}(3)$. We then have minimal weight
(positive energy, which is one of the reasons for choosing AdS) unitary
irreducible representations (UIRs) of a corresponding Lie group. In the
following we consider mainly the twofold covering $SO_{(2,3)}$ of the connected
component of the identity of ${\rm{SO}}(2,3)$, and denote by $D(E_0,s)$
these minimal weight
representations. Here $E_0$ is the minimal ${\rm{SO}}(2)$ eigenvalue and the
half-integer $s$ is the spin. These irreducible representations are
unitary (belonging to the discrete series above the limit of unitarity)
provided $E_0\geq s+1$ for $s\geq 1$ and
$E_0\geq s+\frac{1}{2}$ for $s=0$ and $s=\frac{1}{2}$.

At the limit of unitarity (i.e. when $2E_0$, which is an integer
for $SO_{(2,3)}$ but can take any value for the universal covering,
tends to the limit from above), the Harish Chandra module $D(E_0,s)$
becomes indecomposable and the physical UIR appears as a quotient,
a hall-mark of gauge theories.
For instance, for $s\geq 1$, we get in the limit an indecomposable
representation denoted here by $ID(s)$ or more explicitly by
$D(s+1,s)\rightarrow D(s+2,s-1)$, a shorthand notation \cite{FFS88}
for what mathematicians would write as a short exact sequence of
modules  $0\rightarrow D(s+1,s)\rightarrow ID(s)\rightarrow
D(s+2,s-1)\rightarrow 0$.

Now in gauge theories one needs extensions involving more than two
UIRs. A typical situation is the case of flat space electromagnetism
where one has the classical Gupta-Bleuler triplet which, in our
shorthand notations, can be written $Sc\rightarrow Ph \rightarrow Ga$.
Here $Sc$ (scalar modes) and $Ga$ (gauge modes) are massless
zero-helicity UIRs $h(0,0)$ of the Poincar\'e (inhomogeneous Lorentz)
group $\mathcal{P}_{1+3}= SO_{(1,3)}\cdot\mathbb{R}^4$ while $Ph$ is the
module of physical modes, transforming under $h(0,1)\oplus h(0,-1)$,
where $h(0,s)$ is the UIR of $\mathcal{P}_{1+3}$ with mass 0 and
helicity $s\in \mathbb{Z}$. The scalar modes can be suppressed by a
gauge fixing condition (e.g. the Lorentz condition) but then one is left
with a nontrivial extension $Ph \rightarrow Ga$ on the vector space
$Ph\dot{+} Ga$ which has no invariant nondegenerate metric and cannot
be quantized covariantly. However the above Gupta-Bleuler triplet,
a nontrivial successive extension $Sc\rightarrow (Ph\rightarrow Ga)$,
is an indecomposable representation on a space which admits an invariant
nondegenerate (but indefinite) Hermitian form and it must be used in order
to obtain a covariant quantization of this gauge theory. We shall meet
here a similar situation, which in fact cannot be avoided. Indeed a
general result \cite{Ar85} says in particular that if an extension
$U^1\rightarrow U^2$, with $U^2$ a UIR, has a nondegenerate Hermitian
form, then $U^1$ is equivalent to an extension $U^3\rightarrow U^2$ for
some representation $U^3$ and the original extension is in fact a triplet
$U^2\rightarrow U^3\rightarrow U^2$.

The {\it massless representations} of ${\rm{SO}}(2,3)$
are defined (for $s\geq \frac{1}{2}$) as $D(s+1,s)$  and (for helicity
zero) $D(1,0)\oplus D(2,0)$. There are many justifications to this
definition, among which we can mention \cite{AFFS81}:
\begin{itemize}
\item[a)] The representations $D(s+1,s)$ contract smoothly, in a precise
mathematical sense, to either one of the two massless representations
$h(0,\pm s)$ of $\mathcal{P}_{1+3}$.
Each of the latter has an operationally unique extension to a UIR of
$SO_{(2,4)}$ (a 4-fold covering of the conformal group), the restriction
of which to the $SO_{(2,3)}$ subgroup is exactly the representation we
started with. Moreover each $D(s+1,s)$ can be extended (also uniquely
once the sign is fixed) to either of the two UIRs of the conformal
group which have $h(0,\pm s)$ for restriction to the Poincar\'e group
$\mathcal{P}_{1+3}$. The same properties
are true (for helicity zero) of $D(1,0)\oplus D(2,0)$.
\item[b)] The representations $D(E_0,s)$ can be realized as field
theories on AdS$_4$ but, at the limit of unitarity $D(s+1,s)$ with
$s\geq 1$, they are accompanied by extensions. As a consequence we get
a gauge theory, quantizable only by use of an indefinite metric and
a Gupta-Bleuler triplet.
For $D(s+1,s)$ with $s\geq 0$, the physical signals propagate on
the AdS$_4$ light cone \cite{FFG86}.
\end{itemize}
For $s=0$ and $s=\frac{1}{2}$, the above mentioned gauge theory
appears not at the level of the massless representations
$D(1,0)\oplus D(2,0)$ and $D(\frac{3}{2},\frac{1}{2})$ but at the
limit of unitarity, the singletons Rac$=D(\frac{1}{2},0)$ and
Di$=D(1,\frac{1}{2})$. These UIRs remain irreducible on the Lorentz
subgroup ${\rm{SO}}(1,3)$ and on the (1+2) dimensional Poincar\'e group
$\mathcal{P}_{1+2}$, of which $SO_{(2,3)}$ is the conformal group.
On $\mathcal{P}_{1+2}$ they give \cite{Bi82} the only
massless (discrete helicity) representations.
The singleton representations have a fundamental property:
\begin{equation}\label{di+rac}
({\mathrm{Di}}\oplus{\mathrm{Rac}})\otimes({\mathrm{Di}}
\oplus{\mathrm{Rac}})=(D(1,0)\oplus D(2,0))\oplus
2 \bigoplus_{s=\frac{1}{2}}^\infty D(s+1,s).
\end{equation}
Note that all the representations that appear in the decomposition are
massless representations.
Thus, in contradistinction with flat space, in AdS$_4$, massless states
are ``composed'' of two singletons. The flat space limit of a singleton
is a vacuum (a representation of $\mathcal{P}_{1+3}$ which is trivial on
the translations) and, even in AdS$_4$, the singletons are very poor in
states: their $(E,j)$ diagram has only a single trajectory (hence their
name). The $(E,j)$ spectra of the massless and singleton representations
is:
\smallskip

$D(s+1,s)$, $s>0$: $E-j=1,2,\ldots$; $j-s=0,1,\ldots$\ .

$D(1,0)$: $E-j=1,3,\ldots$; $j=0,1,\ldots$\ ; \
 $D(2,0)$: $E-j=2,4,\ldots$; $j=0,1,\ldots$\ .

Rac$=D(\frac{1}{2},0)$: $E-j=\frac{1}{2}$; $j=0,1,2,\ldots$\ ;\quad
Di$=D(1,\frac{1}{2})$: $E-j=\frac{1}{2}$; $j=\frac{1}{2},
\frac{3}{2}, \ldots$\ .
\smallskip

\noindent [In AdS$_3$, where $\mathfrak{so}(2,3)$ is the conformal
Lie algebra and the anti-De~Sitter Lie algebra is
$\mathfrak{so}(2,2)\equiv\mathfrak{so}(1,2)\oplus\mathfrak{so}(1,2)$,
the ``physical'' massless representations are Di and Rac and the
analogue of singletons are the metaplectic representations
$D(\frac{1}{4})$ and $D(\frac{3}{4})$ of $\mathfrak{so}(1,2)$.
The sum of the two latter is the harmonic oscillator representation
and its tensor square is Di$\oplus$Rac, so we have an exact analogue
of (\ref{di+rac}) \cite{FF80}.
There is however a potentially important difference \cite{HFF92}:
the AdS$_3$ algebra $\mathfrak{so}(2,2)$ is no more the whole
conformal algebra of the $1+1$ space time, since that algebra is
well-known to be infinite dimensional. We shall not elaborate on
this point here.]

In normal units a singleton with angular momentum $j$ has energy
$E=(j+\frac{1}{2})\rho$, where $\rho$ is the curvature of the
AdS$_4$ universe. This means that only a laboratory of cosmic
dimensions can detect a $j$ large enough for $E$ to be measurable
since the cosmological constant (of the order of $\rho$) is very
small. At the flat space limit, the singletons become
vacua (representations of $\mathcal{P}_{1+3}$ with vanishing
energy and momentum) so that they carry no energy at all.
Furthermore local observation of a free singleton field is prevented by
gauge invariance (we shall come back briefly to this point below).
We thus have what can be called ``kinematical confinement'' of singletons
\cite{FF80}, which suggests that they can be a viable alternative
for quarks as fundamental constituents of matter. Elementary particles
would then be composed of two, three or more singletons and/or
anti singletons (the latter being associated with the contragredient
representations). As with quarks, several (three so far) flavors of
singletons (and anti singletons) should eventually be introduced to
account for all elementary particles.
In order to pursue this point further we need to develop a field
theory of singletons and of particles composed of singletons.

\section{Field Theory}
In this section we shall give a very short overview of the many
developments already achieved with singleton field theory and
interactions of singletons.
A first attempt to quantize the singleton field was based on the De Sitter
covariant Klein-Gordon equation ($\square -\frac{5}{4}\rho)\phi=0$ 
where $\rho =\ 3\Lambda,~\Lambda$ the cosmological constant.  
An appropriate choice of boundary conditions, \break 
$\lim r^{1\over 2}\phi <\infty$ as $r\rightarrow\infty$,
leads to a space of solutions that
carries the singleton representation $D(\frac{1}{2},0)$ but not as an
invariant subspace.  Instead, $D(\frac{1}{2},0)$ is induced on a 
quotient space of solutions, where the ignorable invariant
subspace consists of the solutions that satisfy
$\lim r^\frac{1}{2}\phi\rightarrow 0$ as $r\rightarrow\infty.$  This is a
difficulty even in the context of
classical field theory,  for it means that there is no invariant propagator
that includes the contributions from the
singleton modes.  An invariant propagator does exist, but an examination of
its asymptotic properties reveals that all its
Fourier modes fall off as $1/r^{5\over 2}$ at infinity; these modes
constitute an invariant subspace on which the space time symmetry group
acts by $D(\frac{5}{2},0)$.

It is very significant that this representation on the ignorable ``gauge"
modes is of the ordinary massive type, while the singleton representation 
is highly degenerate.  The energy levels of the former are degenerate 
and the spectrum of angular momentum is limited from above by the energy
(in units of $\rho$). The energy levels of $D(\frac {1}{2},0)$ are
much more degenerate: $l=E-\frac{1}{2}$.  This suggests that the physical,
singleton modes are swamped by the gauge
modes and that any interaction designed to detect singletons will fail to
be gauge invariant and hence non-unitary.

In the idiom of representation theory, the space of solutions of the equation
$(\square  -\frac{5}{4}\rho)\phi\ = 0$ satisfying the boundary condition
$r^{1\over 2}\phi\ <\infty$  as $r\rightarrow\infty$, carries the 
non-decomposable representation 
$D(\frac {1}{2}, 0)\rightarrow\ D( \frac {5}{2},0)$.
Quantization needs a non-degenerate, invariant symplectic structure. This
requires the introduction of additional modes, canonically conjugate to 
the gauge modes (compare the situation in electrodynamics where Maxwell 
theory has no momentum conjugate to gauge modes), to give to the total
space the symmetric form 
\begin{equation}
D(\frac {5}{2}, 0)\rightarrow\ D(\frac{1}{2}, 0)\rightarrow\ 
D (\frac {5}{2}, 0)
\end{equation}
or `` scalar $\rightarrow$ transverse $\rightarrow$ gauge". Initially,
this was done by admitting logarithmic solutions of the Klein-Gordon 
equation above. Afterwards it was discovered that the dipole equation
$( \square\ -\frac {5}{4}\phi)\ ^{2}\phi\ = 0$
with the same boundary conditions, provides a much more interesting solution
to the problem.

It is remarkable that this particular instance of the dipole equation, in
marked contrast with what is the case in flat space, and also in 
anti De Sitter space with any other value of the mass
parameter, actually contains physical propagating
modes.  (In all the other cases the representation takes the form 
``scalar$\rightarrow$gauge", with no physical section in between.)  
What is even more remarkable is that this theory is a  {\it topological
field theory}; that is, the physical solutions manifest themselves only 
by their boundary values at $r\rightarrow\infty$: lim $r^{1\over 2}\phi$ 
defines a field on the 3-dimensional boundary at infinity.  There, on the
boundary, gauge invariant interactions are possible and make a 3-dimensional
conformal field theory.  This is a 4-dimensional analogue of the
5-dimensional anti DeSitter/4-dimensional conformal field theory duality
discovered recently by Maldacena \cite{Ma97}.

However, if massless fields (in 4 dimensions) are singleton composites,
then singleton must come to life as four dimensional objects, and this
requires the introduction of unconventional statistics.  The requirement
that the bilinears have the properties of ordinary (massless) bosons also
tells us that the statistics of singletons must be of another sort.

The basic idea is \cite{FF88,FFS88} that we can decompose the
singleton field operator as
$\phi(x)=\sum_{-\infty}^\infty \phi^j(x)a_j$ in terms of positive
energy creation operators $a^{*j}=a_{-j}$ and annihilation
operators $a_j$ (with $j>0$) without so far making any assumptions
about their commutation relations. The choice of commutation relations comes
later, when requiring that photons, considered as two-Rac
fields (using the full tensor product of the two singleton
triplets) be Bose-Einstein quanta. The singletons are then subject
to unconventional statistics \cite{FF89} (which is perfectly
admissible since they are naturally confined) and an appropriate
Fock space can be constructed. Based on these principles, a
(conformally covariant) composite QED theory was constructed
\cite{FF88}, with all the good features of the usual theory.
In addition one can show \cite{FF87} that the BRST structure of singleton
gauge theory induces the BRST structure of the electromagnetic
potential. 
Conformal covariance is based \cite{BFH83} on the
indecomposable $\mathfrak{so}(2,4)$ Gupta-Bleuler triplet
$D(1,\frac{1}{2},\frac{1}{2})\rightarrow
[D(2,1,0)\oplus D(2,0,1)\oplus \mathrm{Id}]\rightarrow
D(1,\frac{1}{2},\frac{1}{2})$
which gives by restriction two inequivalent De Sitter triplets
$D(3,0)\rightarrow D(2,1)\rightarrow D(3,0)$ and
$D(1,1)\rightarrow [D(2,1)\oplus\mathrm{Id}]\rightarrow D(1,1)$,
both of which   appear in the direct product of $D(\frac {5}{2}, 0)
\rightarrow\ D(\frac {1}{2}, 0) \rightarrow\
D(\frac {5}{2}, 0)$ by itself.

This procedure can be (and has in great part been) extended to
the spinor singleton (the Di) and both Di and Rac can be
combined to give a superfield formulation covariant under
the superalgebra $\mathfrak{osp}(4\vert 1)$ \cite{Fr88,FF98}.
This will permit to include Yang-Mills fields, quantum gravity,
supergravity and models of QCD, all based on singletons as
fundamental constituents.

The latest contribution \cite{FF98} to this interpretation of massless 
fields as singleton composites deals with gravitons, giving an explicit 
expression for the weak gravitational potential in terms of singleton 
bilinears. 
If this idea is introduced in the context of bulk/boundary duality, then it
is natural to relate massless fields on the bulk to conserved currents on
the boundary. But we are interested in the composite nature of massless 
fields on space time (the bulk), and a direct current-field identity is
then inappropriate, since currents are conserved by virtue of
the field equations while massless fields are divergenceless only on the
physical subspace defined by gauge fixing. 
In the paper \cite{FF98} it was shown that the dipole formulation
provides a natural construction of all massless fields in terms of
bilinears that are conserved only by virtue of the gauge fixing 
condition on constituent singleton fields.

Now remember that the ``massless'' De Sitter representation
$D(\frac{3}{2},\frac{1}{2})$ (in contrast with other $D(s+1,s)$
representations), has spin above the limit of unitarity, a fact that
singles it out among massless AdS$_4$ particles. It can be obtained
as one of the two, $\gamma_5$-related, irreducible representations that
constitute the space of solutions of the corresponding Dirac equation.
By developing a field theory of composite neutrinos along the lines
explained above (neutrinos composed of singleton pairs, with three flavors
of singletons) it might be possible to {\it correlate} the recently observed
oscillations between the two or three kinds of neutrinos (that
suggests they should have a mass and gives some estimates of it) with
the {\it AdS$_4$ description} of these ``massless" particles. To avoid
misunderstandings we want to stress that we are fully aware of the fact that
any reasonable estimate of \underbar{the value} of the cosmological constant
rules out a direct connection to the value of experimental parameters like
PC violation coupling constants or neutrino masses. (PC violation is a
feature of composite QED, though no estimate of its strength has been 
made.) What we are saying is that \underbar{the structure} of
Anti De Sitter field theory, and more especially the structure of
singleton field theory, may provide a natural framework for a
description of neutrino oscillations.


\noindent {\textbf{PACS (1998)}}: {\small 04.62.+v, 04.65.+e, 11.15.-q,
11.25.Hf, 11.30.Pb, 12.20.-m, 12.60.Rc, 14.80.-j.}

\noindent  {\textbf{MSC (1991)}}: 81T20, 81T70, 81R05; 83E50, 83E30, 17A42,
17A30.
\end{document}